\documentclass{hotnets17}

\usepackage{times}  
\usepackage{hyperref}
\usepackage{graphicx}
\usepackage{authblk}

\hypersetup{pdfstartview=FitH,pdfpagelayout=SinglePage}

\newcommand{\ap}{{\scshape AppSwitch}}

\usepackage{amssymb}
\usepackage{pifont}
\newcommand{\cmark}{\ding{51}}%
\newcommand{\xmark}{\ding{55}}%

\author[1]{Dinesh Subhraveti}
\author[1]{Sri Goli}
\author[1]{Serge Hallyn}
\author[1]{Ravi Chamarthy}
\author[1,2]{Christos Kozyrakis}
\affil[1]{Fermat Inc.}
\affil[2]{Stanford University}

\begin{document}



\title{AppSwitch: Resolving the Application Identity Crisis}

\maketitle

\begin{abstract}

Networked applications traditionally derive their identity from the identity of the host on which they run.  The default application identity acquired from the host results in subtle and substantial problems related to application deployment, discovery and access, especially for modern distributed applications.  A number of mechanisms and workarounds, often quite elaborate, are used to address those problems but they only address them indirectly and incompletely.

This paper presents \ap{}, a novel transport layer network element that decouples applications from underlying network at the system call layer and enables them to be identified independently of the network.  Without requiring changes to existing applications or infrastructure, it removes the cost and complexity associated with operating distributed applications while offering a number of benefits including an efficient implementation of common network functions such as application firewall and load balancer.  Experiments with our implementation show that \ap{} model also effectively removes the performance penalty associated with unnecessary data path processing that is typical in those application environments.

\end{abstract}

\section{Introduction}

The identity of networked applications is traditionally derived from the identity of the host on which they run.  However, applications and infrastructure hosts are very different objects that need to be referenced independently by very different entities.  A web server, for example, is identified by an IP address even though IP addresses were designed to identify network hosts rather than application endpoints.

The practice is not unacceptable as long as a relatively stable one-to-one mapping can be maintained between the application and its host.  That was the case for early networked applications which were undistributed, deployed on a dedicated host and left undisturbed for a long time.  Modern applications, which are distributed, mobile, ephemeral and run on multi-tenant infrastructure, no longer conform to that simplistic model.  It is especially the case for microservices applications~\cite{microservices} where otherwise monolithic software is built and deployed as a distributed set of sub-applications called microservices.  Individual microservices move about from host to host, are often short-lived and share an underlying pool of stateless infrastructure hosts with other distributed applications and their microservices.

The practice of associating applications with their hosts is so common and ubiquitous that it is hardly questioned. However, something as fundamental as application identity naturally has a bearing on almost every aspect of application connectivity and networking.  In particular, improperly defining and identifying applications leads to subtle and substantial problems related to common network functions with which applications have to directly interface.  While lower level network functions such as packet switching do not depend on applications, network functions such as application discovery, segmentation, firewalls, load balancers, VPN gateways, implicitly depend on well-defined application identities.  A number of mechanisms~\cite{k8s,dockernet,dns,envoy} and elaborate devices have been built over the years to work around the symptoms of application misidentification.  The workarounds only address the issues indirectly and incompletely and typically produce new problems.

We present \ap{}, a novel transport layer network element that decouples applications from the network and forms a clear interface between the two.  Like the way a router provides IP connectivity at the IP layer and the way a bridge provides L2 connectivity at the link layer, \ap{} operates at the transport layer and provides mutual discovery and connectivity to applications.

\ap{} uses the system call interface to decouple applications from the network.  As a well-defined and stable interface and the exclusive means for applications to access the network and its properties, it provides a convenient and powerful layer to decouple and virtualize application's interactions with the network.  In contrast to low level approaches like overlays~\cite{vxlan} that have to process every packet or high level approaches like proxies~\cite{envoy} that have to move data between connections, virtualization of the network API avoids data path processing.

The decoupled interface enables applications to be identified independently of the network such that typical networking issues with modern distributed applications are removed.  In addition, it provides a number of other advantages including reduced operational cost and complexity by minimizing unnecessary friction between applications and operations teams, effective and efficient implementation of application-level network functions, ability to run applications across heterogeneous infrastructure backends including bare metal machines, VMs, containers and cloud, and improved performance by selecting most suitable network medium.  \ap{} achieves these benefits without requiring any changes to the applications or the infrastructure.

This paper focuses on the design and implementation of \ap{}.  Section 2 exposes the problems due to coupling between applications and network.  Section 3 enumerates the properties required of a desirable solution and analyzes existing approaches and related work in that light.  Sections 4 presents \ap{}'s high level model.  Section 5 discusses the details of its architecture and key components.  Section 6 provides preliminary results of our experiments.

\section{Coupling Between Applications and the Network}

Even though traditional network stack~\cite{osi} is based on well-defined layers that separate application and network level functionality, applications are closely tied to network level artifacts such as IP addresses.  Likewise, network functions depend on application level constructs which they often infer through techniques such as deep packet inspection.  While the architecture is built for interoperability between layers, it is not intended to support the behavior of modern distributed applications which requires a clean separation between applications and underlying network to enable them to run across hosts.

The unintended coupling between applications and infrastructure has a significant impact on the cost of operating complex distributed applications which are becoming the necessary drivers for almost every modern enterprise.  These organizations typically consist of applications and operations counterparts each responsible for building and operating the applications respectively.  The productivity and efficiency of those teams depend on minimizing the interactions between the two counterparts.  While the applications team can independently make decisions about the internals of the application, they would have to depend on the operations team to assign identities to their application endpoints.  For example, they could independently choose to serialize application data structures in JSON format over an HTTP channel between the application instances but they cannot choose the names or IP addresses of those instances.

In order to deploy an otherwise self-contained distributed application, the applications team would have to acquire a set of IP addresses from a shared pool which is arbitrated by the operations team.  An obvious consequence of the default identity that the applications acquire from their hosts is that any change to the application's host or its identity invalidates previously advertised references to the application.  A host reboot, expiration of DHCP lease, rescheduling of the application instance to a different host or live migration of the underlying VM to a different subnet etc. could all cause the application to become unreachable at its former identity.  Any policy specification that references application endpoints through their host identities would become invalid as application's identity changes.  For example, any firewall rules meant to segregate application endpoints established within the network infrastructure would become invalid.  Routers and switches typically expose proprietary interfaces to update their configuration and it is often a tedious and error-prone task to update it.

The operational burden is further compounded by the need for simpler policy specifications and route tables based on route summarizations that require IP addresses to be assigned according to the physical topology of the datacenter racks.  That in turn restricts the applications to specific nodes or racks.  The problem is particularly pronounced in hybrid and cloud based application environments where allocation of network resources is regulated by the cloud providers.  In case an application needs to be moved to a different host or a region, the operations team would have to reassign the IP addresses and update firewall rules accordingly.  In some cases such a reassignment may not be possible.

\section {Existing Approaches and Related Work}

Any approach that can effectively decouple applications from the network and provide them distinct identity must meet three important requirements.  First, it must support the behavior of modern distributed applications and their operating environments.  From first principles, application identifiers provided by such an approach must be unique and consistent.  That is, an identifier must uniquely identify the application and remain constant during its lifetime.

Second, the approach should support existing applications and existing network infrastructure.  Rewriting or relinking applications to use a new RPC library or a service discovery protocol is typically not practical.  Reconfiguring applications to use alternate names or ports is also a nontrivial task.  The approach also should not place unusual requirements or load on the network infrastructure.  Legacy networking environments are rigid and proprietary.  Upgrading them to support a new facility or increasing their resources (such as TCAM table space), or deploying and maintaining new infrastructure components is typically too expensive.

Third, the approach must not introduce unacceptable operational or computational cost.  Particularly, the system should be simple to deploy and operate.  Operational cost and complexity of traditional application environments is one of the key challenges to be addressed and the solution should not introduce new operational burden of its own.  The system also should not adversely affect application or network performance.

The general practice of using a combination of IP address and port number does not meet these requirements.  IP addresses are unique in a network but they cannot serve as consistent application identifiers because they cannot remain constant as the application moves from host to host.  Port numbers on the other hand can be consistent because the same port number may be available on every host.  However, they cannot independently serve as application identifiers given that the scope of their uniqueness is limited to the host and they are always tied to IP addresses.  Standardization of well-known ports~\cite{stdports} further limits their utility as identifiers with its implicit assumption that no two instances of the same application are run on the same host.  For example, the same host cannot run two web servers if they were both to use the default HTTP port.  This is a common occurrence in environments with shared infrastructure and in modern REST applications where every application acts as a web server providing its services to other applications over HTTP.

Several techniques~\cite{k8s,dockernet,dns,envoy} are used to offset the inherent shortcomings of IP addresses and port numbers to act as application identifiers but none of them is able to provide both uniqueness and consistency required to support distributed applications.  Existing systems tend to use a combination of techniques to support the requirements of distributed applications.  They involve assigning each application either a unique IP address or a unique port and then using a layer of indirection to keep them consistent.  As an example, Docker~\cite{dockernet} assigns each application container a different IP address and uses an underlying overlay network to keep them from changing as the application moves across hosts.  Envoy~\cite{envoy} on the other hand assigns each application service a unique port number and factors out the IP address component of its identity by having all client applications go through a local proxy running on the loopback address.  The proxy then directs the client requests based on a mapping between the unique port number and the set of backend servers providing the service.

\begin{table}[h!]
\label{tab:comparison}
\begin{tabular}{|c|c|c|c|c|c|c|}
 \hline
 			& \multicolumn{2}{|c|}{Dist App} & \multicolumn{2}{|c|}{Transparency} & \multicolumn{2}{|c|}{Overhead} \\
 \hline
			& Uniq & Const & App & Infra & Comp & Ops \\
 \hline
 \hline
 PAIP & \cmark & \xmark & \cmark & \xmark & \cmark & \xmark \\
 \hline
 PAP & \cmark & \xmark & \xmark & \cmark & \cmark & \xmark \\
 \hline
 OVR & \xmark & \cmark & \cmark & \xmark & \xmark & \xmark \\
 \hline
 NAT & \xmark & \cmark & \cmark & \cmark & \xmark & \xmark \\
 \hline
 DNS & \xmark & \cmark & \xmark & \xmark & \cmark & \xmark \\
 \hline
 PXY & \xmark & \cmark & \xmark & \cmark & \xmark & \xmark \\
 \hline
 \hline
 ASW & \cmark & \cmark & \cmark & \cmark & \cmark & \cmark \\
 \hline
\end{tabular}
\caption{Comparison of approaches that address identity of distributed applications}

\end{table}

Table 1 loosely compares \ap{} (ASW) with the capabilities of the techniques typically used to ensure uniqueness and consistency of application identifiers.  Per-application IP addresses (PAIP) and per-application ports (PAP) provide uniqueness and overlays (OVR), dynamic DNS (DNS), network address translation based on IPtables (NAT) and application proxy (PXY) provide consistency.  Per-application IP increase the burden on the network and the cost of operating it.  Calico~\cite{calico} for example treats each application container as a network host in its own right and assigns it a routable IP address.  Using NAT to multiplex applications through the same interface would not solve the problem because assigning the same IP address to multiple applications would prevent them from being scheduled to different hosts without changing their identity.

Assigning per-application port numbers involves reconfiguring applications to bind to nonstandard ports.  Every client application also has to reconfigured to find its services at their custom ports.  Keeping track of port numbers used by every service in a complex distributed application and ensuring their consistent use can be immensely challenging.

Network overlays typically require an elaborate deployment and maintenance effort.  Some of the implementations depend on specific infrastructure capabilities.  Dynamic DNS approach cannot support applications that directly use IP addresses and may also impact correctness due to stale records.  Reducing TTL may avoid stale entries but that would impact performance.  IPtables based address translation suffers from scalability constraints due to the large number of rules that typically need to be installed.  Proxy based approaches are not application transparent and require additional infrastructure and data path processing.

Several recent approaches~\cite{calico,cilium} directly address networking for microservice applications, typically running as containers.  They all require applications to be rewritten or modified in some way.  Consul~\cite{consul} provides a central registry of services that arbitrates discovery between clients and servers.  Servers are required to advertise themselves by registering with the registry which clients would query to discover the location of the servers they want to reach.  While the approach is sound, it doesn't help with existing applications.  A similar mechanism is implemented at the networking layer as an extension to DNS~\cite{dns} that provides new type of records that map names to service endpoints.  It also requires applications to be rewritten to utilize the new facility.  FreeFlow~\cite{freeflow} addresses the performance of container networking by choosing the most optimal network medium.  However, it doesn't support existing applications and relies on container-specific frameworks.

\section {AppSwitch Model}

\ap{} is designed to be a transport layer network element that serves as an interface between applications and the network.  A set of \ap{} instances, running one per host, form an {\it application network} that provides seamless discovery and connectivity to the applications.  Applications on a host are added to the \ap{} instance on the host to join the application network.  \ap{} transparently tracks the applications' execution to detect server applications and their service ports as they come up and propagates their location information across the cluster over a gossip protocol~\cite{swim}.  When a client attempts to reach a server, the request is directed to the appropriate server endpoint.  The details of the mechanisms used to track applications and to propagate location information are discussed in the next section.

Decoupling applications from the network allows them to be named independently of network constraints.  When an application is brought up, the user optionally assigns it a virtual IP address, completely independently of underlying network, simply by specifying it as a parameter to \ap{} when the application is added.  The virtual IP address serves as a unique and consistent identifier representing the application.  Even though the identifier takes an IP address format, it bears no relation to the IP addresses carried by network hosts.  The choice is only driven by backward compatibility with existing applications that expect an IP address.

The user may also specify application identity as a DNS name rather than an IP address, in which case \ap{} assigns it an internal IP address.  A built-in DNS server overrides client application's DNS lookups to return the internal IP address.  When the client in turn asks to connect to that IP address, its request is transparently directed to the server represented by the name.

In addition to a name, the user may also qualify the application's identity with a set of tags in the form of key-value pairs which are propagated along with the location information.  Tags could represent attributes such as security groups that the application belongs to.  When a client tries to reach a service, respective tags are consulted for a match before the connection is allowed.  Connection requests to \ap{}-managed applications from unidentified clients are disallowed.  While a simple grouping may suffice in most cases, a more expressive policy could be easily supported.

Some of the \ap{} instances that have access to external network are designated as {\it gateways} to allow internal server applications to be reached from external clients and external server applications to be reached from internal clients.  When an application is added to \ap{}, the user can optionally indicate that the application is exposed to the external world, in which case, the service is made available through a specific port on one of the gateways.  When an external client connects to the gateway host on the external interface, \ap{} would proxy the connection to corresponding server within the application network.  Typically only a small fraction of the applications of a distributed application require exposure through \ap{} proxy.

By empowering the user to directly specify application identities, \ap{} removes the operational friction of acquiring IP addresses and names.  While network level identifiers referenced by intermediate network infrastructure can be machine-generated, the responsibility of assigning a meaningful name to a high-level application service ultimately rests with a human user, typically in the role of an application or a network architect.  In that role, the user would have a global view of the broader distributed application and its environment and would be able to ensure that unique names are assigned to disjoint applications.

If the user does not name the application, it would not carry any externally referenceable identifier and would not be able to act as a server application.  It can still be a client application that accesses other servers.  This is in contrast to traditional application addressing where both client and server applications carry the network identity of the host.  While client and server network endpoints both need network level identifiers for the packets to flow between them, only server applications need to be identifiable and referenceable at the application level.

If the user does name an otherwise client application, the specified IP address is conveyed as client's identity to the servers the application connects to.  If no such IP address is provided, \ap{} internally assigns and uses a link-local address~\cite{linklocal} for that purpose.  If the same server application exposes multiple services by binding to multiple ports, they would all be individually referenceable through the application's IP address and respective port number of the service.

If the user specifies the same IP address to multiple applications, they would be treated as being a part of the same {\it distributed application} that provides a common port namespace stretching across those applications.  A service exposed by an individual application of a distributed application can be referenced with the name of the distributed application and the service port number, regardless of the specific application that exposes it or the specific host where it is running.  Individual applications could access each other simply through the loopback address as if they are running on the same host. This allows a distributed microservices application to be developed and tested on a single node over loopback interface and then deployed to production as a scalable distributed application without further reconfiguration and without concern about the identities of its endpoints.  \ap{} removes the added complexity of microservice applications compared to their monolithic counterparts by providing a simple and unified virtual host view.

If multiple applications of a distributed application bind to the same port, they would be treated as instances of the same load balanced service.  Requests to connect to that port would be served from one of the available server instances.  Normally, binding to a port which is already in use results in port conflict error.  Instead of flagging it as an error, \ap{} uses it as a simple and intuitive interface to provide a distributed load balancer.  Multiple instances of the same application can be simply brought up to stitch them into a cohesive load balanced service without added infrastructure cost and complexity associated with traditional load balancers.  Because \ap{} implements the logic of selecting the server instance on the client-side, it is more scalable and because there is no proxy involved, it is more efficient.

\begin{figure}[h!]
\includegraphics[width=3.25in,height=1.25in]{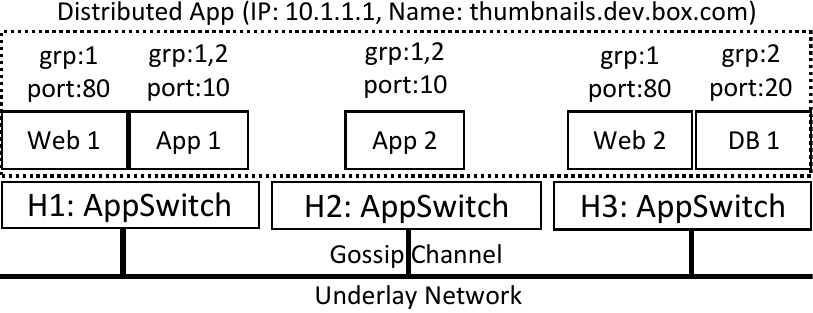}
\caption{\ap{} Model}
\label{fig:model}
\end{figure}

\ap{} model provides for a simple and intuitive specification of distributed microservice applications.  Several application specification formats~\cite{terraform,compose} are being proposed to represent the makeup of distributed applications and deploy them in an infrastructure-agnostic fashion.  However they tend to be excessively complex due to unclear boundary between applications and the network.  \ap{} defines a distributed application with a consistent identity uniformly shared by its applications.  Figure \ref{fig:model} shows a sample deployment of a distributed application on three hosts, {\tt H1}, {\tt H2} and {\tt H3}, consisting of two instances of {\tt Web}, two instances of {\tt App} and one instance of {\tt DB}.  Services provided by these applications are consistently identified by port numbers from a global namespace spanning the distributed application.  Both load balanced instances of {\tt Web} and {\tt App} are consistently represented by their same respective ports even though they belong to the same distributed application.  Two groups, 1 and 2, respectively consisting of {\tt Web} and {\tt App}, and {\tt App} and {\tt DB} are defined by attaching {\tt grp} tags to the applications such that {\tt Web} applications cannot directly talk to {\tt DB}.  These simple constructs enable complex application environments to be constructed through a hierarchical composition of cohesive distributed application units that can systematically reference services exposed by each other.

\section {AppSwitch Architecture}

\begin{figure}[h!]
\includegraphics[width=3.25in,height=2.5in]{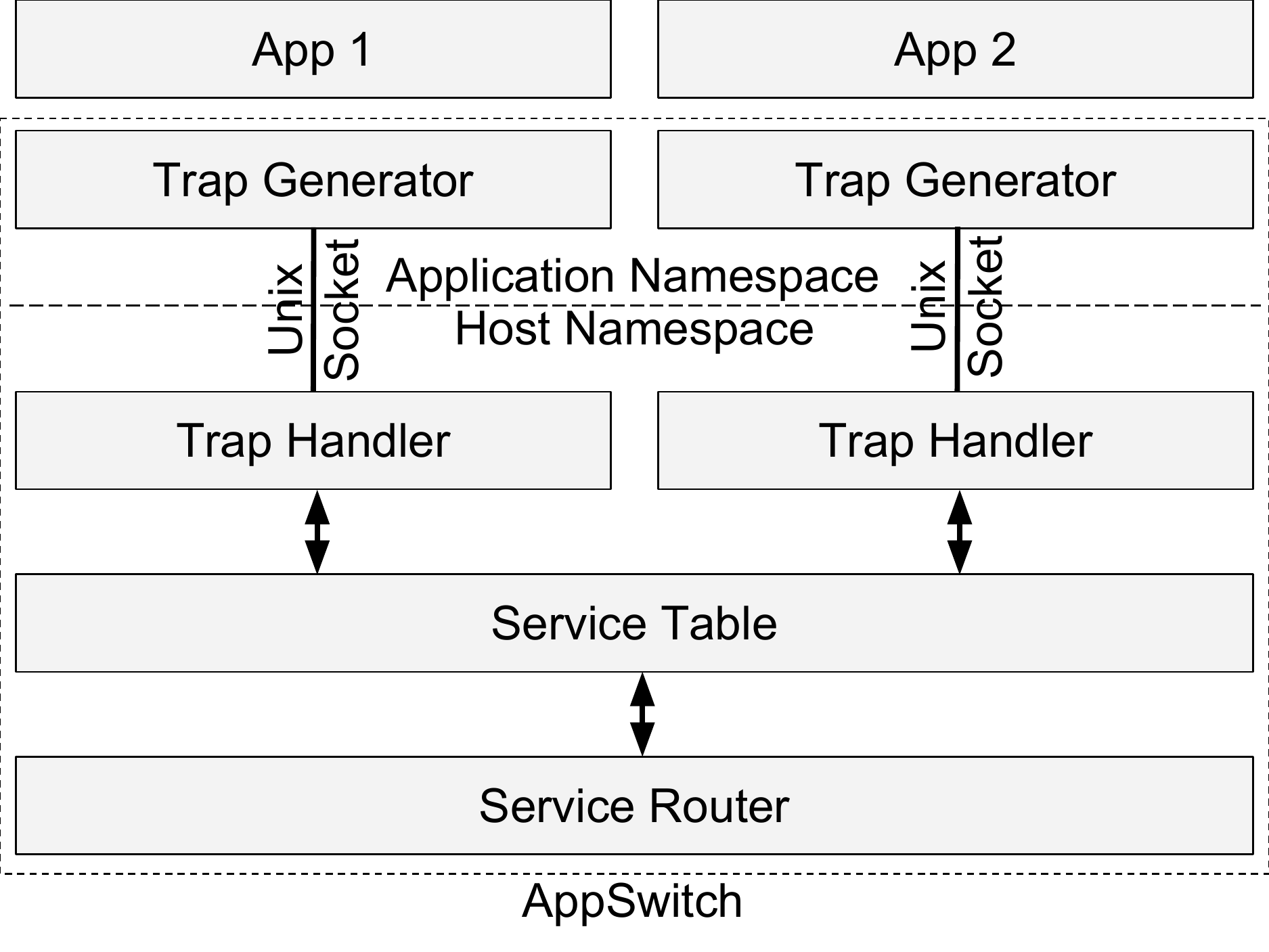}
\caption{\ap{} architecture}
\label{fig:arch}
\end{figure}

Figure \ref{fig:arch} shows the architecture of \ap{}.  It consists of two key components namely {\it trap mechanism} and {\it service router} which share a data structure called {\it service table} that maintains a mapping between application identifiers and network level identifiers.  Service router efficiently propagates the contents of service table with other instances of \ap{} on other hosts over a gossip protocol~\cite{swim}.  The rest of this section describes the trap mechanism.

\subsection {Trap Mechanism}

Trap mechanism provides transparent application instrumentation.  It is conceptually similar to FUSE~\cite{fuse}.  Whereas FUSE enables file systems to be implemented in user space by redirecting an application's file system calls to a user space handler, trap mechanism redirects the network primitives of an application to a user space handler by interposing its network-related system call.  In contrast to FUSE, however, only control plane system calls that carry endpoint identities as parameters are intercepted but not system calls such as {\tt read} and {\tt write}.

Trap mechanism is implemented by two components: {\it trap generator} and {\it trap handler}.  Trap generator intercepts network control plane system calls of the application and forwards them to trap handler over a Unix domain socket.  Trap handler services them and returns their result back to trap generator which in turn conveys it to the application.

When an application is first added to \ap{}, a new network namespace is created to host the application.  An instance of trap generator and trap handler are associated with the namespace such that relevant system calls of the application are intercepted by the trap generator within the application's network namespace and serviced by trap handler running in the host network namespace.  Together, trap generator and trap handler extend host's network connectivity into application's network namespace.  Given that the network namespace is deliberately left empty without any network devices, the trap mechanism forms the only means of network access for the application.  It ensures that every network access is arbitrated by \ap{}.  Network namespace also provides a convenient abstraction that clearly defines what constitutes an application endpoint and a boundary at which application's system calls are intercepted.

Trap generator and trap handler are connected to each other through a Unix domain socket which allows active file descriptors to be passed between them in addition to any static data.  In particular, trap handler can return active file descriptors created in the host network namespace into the application's namespace via trap generator.  When trap generator forwards {\tt socket} system call made by the application, for example, trap handler would create the socket in the host namespace and pass its reference to the application.

In general, trap handler performs appropriate security checks, matches relevant tags associated with the application and negotiates socket connections with right application endpoints on behalf of the application before returning them to the application.  When a client application attempts to connect to a server at the IP address specified by the user during server creation, trap handler looks up the IP address passed by the application in the service table to find the IP address where the server is actually listening.  It then establishes a connection with the server at its real IP address and returns the connected socket to the application.

Once the connection is fully established and returned to the application, it would simply use it as a bitpipe without regard to the endpoint identities of the connection.  Application would not care about the protocol either as long as  a file descriptor abstraction is supported for IO.  In fact, \ap{} returns a Unix socket rather than a TCP socket in case client and server happen to be on the same host.  With additional virtualization, other types of communication media~\cite{freeflow} or low level IO acceleration techniques ~\cite{dpdk} could be used as well.  If the application queries the identities through system calls like {\tt getsockname} or {\tt getpeername}, consistent responses expected by the application are presented.  Likewise, trap handler stays out of the data path once the connected socket is passed to the application.  In case of datagram protocols like UDP, trap handler also services system calls like {\tt sendmsg} and {\tt rcvmsg} that carry endpoint identities in their parameters.

When a server application attempts to listen on an IP address and a port, trap handler binds to any available host interface and any available port on behalf of the application and adds an entry to the service table that maps the incoming IP address and port to the real IP address and port.  The newly added entry is then advertised among other instances of \ap{} running on other hosts by the service router.

\section {Preliminary Results}

We have implemented \ap{} on Linux.  Ease of deployment and operation was one of the primary factors that drove our implementation.  To that extent, \ap{} is built as a simple RPM consisting of a kernel module that implements the trap generator and a statically linked user space utility that implements all other components.  Deploying \ap{} involves installing the RPM and providing only one piece of configuration that points the \ap{} instance to one of the existing instances, if any, to form the cluster that supports the gossip channel.  No other configuration or change is required to the applications or infrastructure.  We also implemented a version of \ap{} using Linux ptrace primitives that doesn't require a kernel module.  The mechanism is conceptually similar to the kernel version.  The preliminary results we present here are based on the user space implementation.

\begin{figure}[h!]
\includegraphics[width=3.25in,height=2.5in]{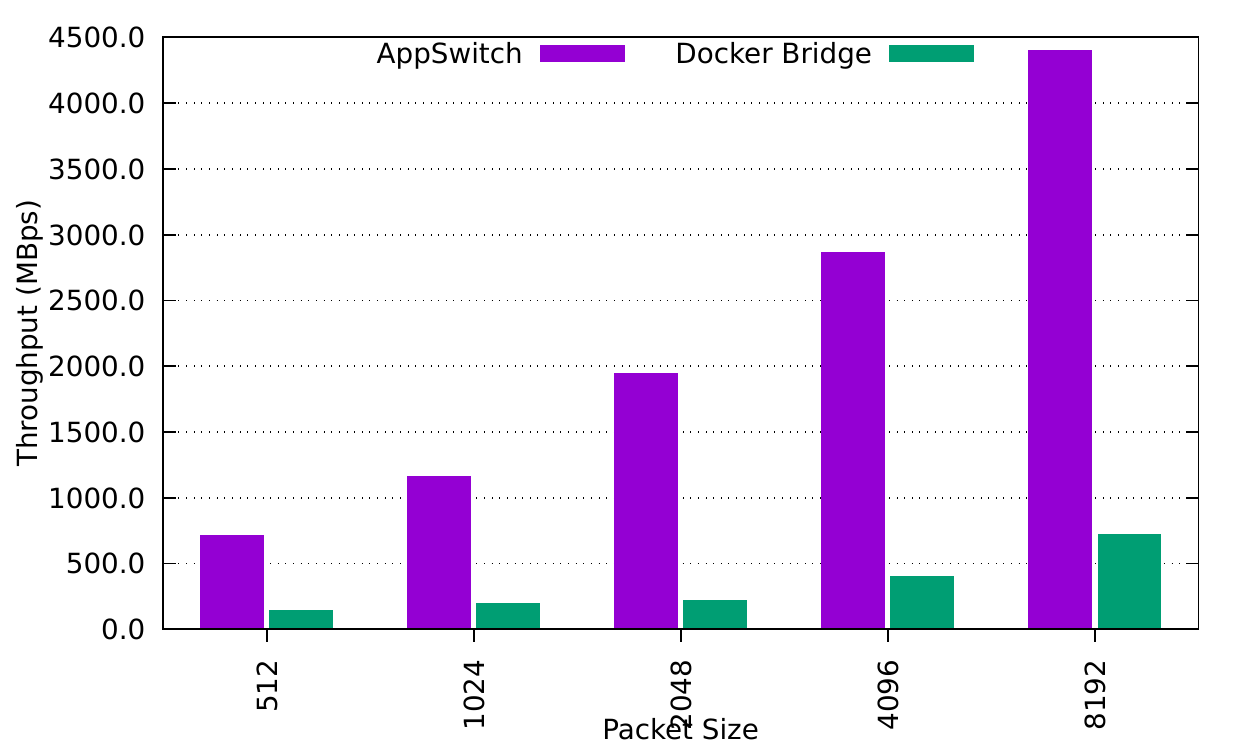}
\caption{Network throughput with \ap{}}
\label{fig:perf}
\end{figure}

The experiments were conducted on a bare-metal machine with Intel Xeon CPU E5-2660 v3 2.60GHz running Centos 7.  We measured the throughput between the client and server instances of {\tt iperf-2.0.9}, each running in a Docker container on the same machine.  The experiment was conducted with the standard Docker network configuration that uses a Linux bridge for inter-container connectivity and then repeated with \ap{} as the network backend.  In case of \ap{}, {\tt iperf} command was prefixed with {\tt appswitch} command along with its options.  Particularly, {\tt iperf} server was given an IP address, a name and a tag that places it in a new group by passing those three options as arguments to {\tt appswitch}.  Client was also brought up with {\tt appswitch} command but without a name or IP address but with a tag that places it in the same group as the server.  The name assigned to the server was directly passed as an argument to {\tt iperf} for it to connect to the server.  By default Docker created virtual interfaces in its containers and connected them to the host network through a Linux bridge.  But those interfaces were ignored in the case of \ap{} and the data only flowed through the channel setup by \ap{}.

The results of the experiment are shown in Figure \ref{fig:perf}.  Throughput was generally higher at larger packet sizes as expected and it was several times higher with \ap{} compared to Linux bridge.  Given that both server and client were running on the same host, \ap{} transparently connected them over a Unix socket even though {\tt iperf} requested a TCP connection.  With Linux bridge, packets had to make two hairpin traversals through the network stack even though both endpoints of the connection were colocated.

\section {Conclusion}

We presented the design of \ap{}, a novel transport level network element that removes the cost and complexity of operating modern distributed applications by effectively decoupling them from the underlying network at the system call layer.  We recognize that the default identity of the applications acquired from the hosts on which they run is the root cause of several subtle and substantial problems and that system call layer provides a convenient and efficient interposition point to address them.  We also show that application-level network functions such as segmentation and load balancing can be implemented more efficiently without incurring the data path processing cost typical of traditional approaches.

\section {Acknowledgements}

As one of the first customers of AppSwitch, Ashar Rizqi and his team provided valuable guidance on various application scenarios that AppSwitch must support at Box Inc.

\bibliographystyle{abbrv} 
\begin{small}
\bibliography{hotnets17}
\end{small}

\end{document}